\documentclass[prb,
twocolumn,
superscriptaddress,showpacs,amsmath,amssymb]{revtex4}

\begin{document}

\title{Dimensionless physics}

\author{G.E.~Volovik}
\affiliation{Low Temperature Laboratory, Aalto University,  P.O. Box 15100, FI-00076 Aalto, Finland}
\affiliation{Landau Institute for Theoretical Physics, acad. Semyonov av., 1a, 142432,
Chernogolovka, Russia}

\date{\today}

\begin{abstract}
{
We discuss two scenarios of emergent gravity. In one of them the quantum vacuum is considered as superplastic crystal, and the effective gravity describes the dynamical elastic deformations of this crystal. In the other one the gravitational tetrads emerge as the bilinear form of the fermionic fields. In spite of the essentially different mechanisms of emergent gravity, these two scenarios have one important common property:   the metric field has dimension of  the inverse square of length $[g_{\mu\nu}]=1/[l]^2$, as distinct from the conventional dimensionless metric, $[g_{\mu\nu}]=1$, in general relativity. As a result the physical quantities, which obey diffeomorphism invariance,  become dimensionless. This takes place for such quantities as Newton constant, the scalar curvature, the cosmological constant, particle masses, fermionic and scalar bosonic fields, etc. This may suggest that the dimensionless physics can be the natural consequence of the diffeomorphism invariance, and thus can be the general property of any gravity, which emerges in the quantum vacuum.
One of the nontrivial consequences of the shift of dimensions is related to topology.  Due to the shift of dimensions some operators become topological, and contain the integer or fractional prefactors in the action. This in particular concerns the intrinsic 3+1 quantum Hall effect and the Nieh-Yan quantum anomaly in terms of torsion.
}
\end{abstract}
\pacs{
}

\maketitle 
 
 \section{Introduction}

It is not excluded that the Standard Model of particle physics together with general relativity are effective theories which emerge at low energy for the collective modes of the extreme ultraviolet.\cite{Bass2020} Sakharov gravity\cite{Sakharov}, which emerges in the fermionic vacuum,  provides the characteristic example. The scenario, where all the known symmetries in our Universe emerge on the macroscopic scale, but disappear in the highly trans-Planckian microscopic regime, takes place in different many-body condensed matter systems. For example, the analogs of Lorentz invariance and the curved spacetime are developed for some low energy fermionic and bosonic modes,\cite{Unruh1981} but these phenomena disappear at large energy, where the microscopic degrees of freedom intervene (the analog of the trans-Planckian degrees of freedom). The condensed matter systems with topologically stable Weyl points in the fermionic spectrum, such as Weyl semimetals and  the chiral superfluid phase of liquid $^3$He, demonstrate simultaneous emergence of chiral fermions, gauge bosons and tetrad gravity,\cite{Nielsen1976,Volovik1986,Froggatt1991,Horava2005,Volovik2003} which do not survive on the high-energy atomic level.

We do not know the structure of the trans-Planckian world, but we can try different possible scenarios of emergent physics and search for the common properties in the low energy corner. Here we consider two scenarios of emergent gravity, which are very different, but have the important common property. 

The first one is the tetrad gravity where the tetrad fields emerge as bilinear combinations of the fermionic fields by symmetry breaking. This scenario has been investigated by Diakonov \cite{Diakonov2011}, Vladimirov and Diakonov\cite{VladimirovDiakonov2012,VladimirovDiakonov2014} and   Obukhov and Hehl.\cite{ObukhovHehl2012}
The analog of the   Diakonov-Vladimirov (DV) scenario takes place in another topological phase of superfluid $^3$He -- the B-phase.\cite{Volovik1990} 

The other one is the analog of gravity in the elasticity theory of crystals,\cite{Bilby1955,Bilby1956,Kroener1960,DzyalVol1980,VolovikDotsenko1979,AndreevKagan84,KleinertZaanen2004,HehlObukhov2007} 
where the elastic deformation are described in terms of the tetrads of elasticity.\cite{DzyalVol1980}  In principle, this analogy can be extended to the real gravity, if the quantum vacuum is considered as a plastic (malleable) fermionic crystalline medium, and the elasticity tetrads become the gravitational tetrads. \cite{KlinkhamerVolovik2019,Zubkov2019} The condensed matter analog of such vacuum is the quantum crystal with fermionic quasiparticles, such as vacancies.\cite{Andreev1969,Dzyaloshinskii1972,Dzyaloshinskii1972b}

The common property of these two approaches to quantum gravity is that  the tetrad fields in both theories have dimension of inverse length. As a result most of the physical quantities which obey diffeomorphism invariance become dimensionless. \cite{VladimirovDiakonov2014,NissinenVolovik2018,NissinenVolovik2019,Volovik2020} Since the two very different scenarios lead to the same phenomenon, it is natural to suggest that the gravity in our universe also follows this common rule. Here we consider some consequences which come from this rule.

\section{Elasticity tetrads}

Let us consider theory of the crystal elasticity using approach of Ref. \cite{DzyalVol1980}. The deformed crystal structure can be described as a system of three crystallographic surfaces of constant phase $X^a(x)=2\pi n^a$, $n^a \in {Z}$ with $a=1,2,3$. The intersection of the surfaces
\begin{equation}
X^1({\bf r},t)=2\pi n^1 \,\,, \,\,  X^2({\bf r},t)=2\pi n^2 \,\,, \,\, X^3({\bf r},t)=2\pi n^3 \,,
\label{points}
\end{equation}
are the nodes of the deformed crystal lattice. For the undeformed crystal, $X^a(\bf{r},t) = \bf{K}^a \cdot \bf{r}$, where $\bf{K}^a$ are the (primitive) reciprocal lattice vectors. The deformations of the crystal can be described in terms of the elasticity tetrads, the  gradients of the phase functions:
\begin{equation}
E^{~a}_i(x)= \partial_i X^a(x)\,.
\label{reciprocal}
\end{equation}
In the absence of dislocations, $E^{~a}_i(x)$ is an exact differential:
\begin{equation}
\partial_k E^{~a}_l(x)-\partial_l E^{~a}_k(x)=0\,.
\label{integrability}
\end{equation} 
In the presence of the  topological defects -- dislocations, the density of dislocations plays the role of torsion:
\begin{equation}
T^a_{kl} =(\partial_k E^{~a}_l-\partial_l E^{~a}_k)\,.
\label{Torsion}
\end{equation}

Such construction can be extended to the model of the 3+1 quantum vacuum with $a=0,1,2,3$, assuming that the vacuum looks like the plastic crystalline medium. In this model the elasticity tetrads $E^{~a}_\mu$ become the gravitational tetrads.\cite{KlinkhamerVolovik2019,Zubkov2019} The deformed vacuum crystal with dislocations describes the curved geometry of the teleparallel Weitzenb\"ock gravity  with vanishing curvature and nonzero torsion.  On the macroscopic coarse grained level, where the separate dislocations are not resolved, the torsion field $T_{\mu\nu}^a$ can be considered as a continuous function of coordinates. 
The metric $g_{\mu\nu}$, which originates from the elasticity tetrads, is:
\begin{equation}
g_{\mu\nu}=\eta_{ab}E^{~a}_\mu E^{~b}_\nu\,,
\label{Metric}
\end{equation}
where $\eta_{ab}=(-,+,+,+)$. 

The important property of the elasticity tetrads is that being the derivatives, they have the canonical dimensions of  inverse length, 
$[E^{~a}_\mu]=[l]^{-1}$. Correspondingly the metric in Eq.(\ref{Metric}) has dimension $[g_{\mu\nu}]=1/[l]^2$. This means that the distance or interval $d s$ in
\begin{equation}
ds^2=-g_{\mu\nu}dx^\mu dx^\nu\,,
\label{Interval}
\end{equation}
 is dimensionless, $[\Delta s]=1$. This simply means that the distance between the two nodes of the deformed crystal is determined by the integer number of crystal surfaces between the points of the grid, and thus does not depend on the length scale. 
 This is natural for such quantum vacua, in which the size of the unit
cell is not fixed and can be arbitrary.  
 
Let us introduce notation $d$ for conventional dimension of the physical quantities, and the notation $d_{\rm DV}$ for the shifted dimension of the same quantities. The shift of dimensions means that in the DV approach    $[l]^{-d} \rightarrow [l]^{-d_{\rm DV}}$. For example, for the interval $ds$ we have the conventional dimension $d=-1$ and the shifted dimension $d_{\rm DV}=0$, see Eq.(\ref{GRdimensions}). For the torsion field one has the conventional dimension $d=1$, while the shifted dimension is $d_{\rm DV}=2$. 

The shift of the dimensions of the physical quantities due to the dimensionless interval leads to the new properties of the  quantum vacua and topological insulators. In particular, this allows us to extend the application of the topological anomalies. For example, the Chern-Simons term describing the 3+1 intrinsic quantum Hall effect becomes dimensionless. As a result, the prefactor of this term is given by the integer momentum-space topological invariants in the same manner as in the case  of 2+1 dimension.\cite{NissinenVolovik2018,NissinenVolovik2019,Volovik2020} The shift of dimension is also important for the Nieh-Yan anomaly.\cite{Volovik2020}

\section{ Tetrads as bilinear combinations of fermion fields}:

In the DV theory\cite{Diakonov2011,VladimirovDiakonov2012,VladimirovDiakonov2014,ObukhovHehl2012} the tetrads are composite fields, which emerge as the bilinear combinations of the fermionic fields. The tetrads appear as the order parameter of the symmetry breaking phase transition (see also Ref.\cite{Akama1978}):
\begin{equation}
e^a_\mu=i \left<\psi^\dagger \Gamma^a \nabla_\mu \psi + \nabla_\mu\psi^\dagger \Gamma^a  \psi \right>
\,.
\label{bilinear}
\end{equation}
The corresponding symmetry breaking scheme is $L_{L}\times L_{S}\rightarrow L_{L+S}$, where $L_{L}$ and $L_{S}$ are two separate  symmetries under Lorentz rotations of the coordinate and spin space respectively. These two symmetries are broken to the diagonal subgroup -- the Lorentz group of the combined  rotations in two spaces,  $L_{L+S}$. In addition this order parameter breaks the $PT$ symmetry, see also Ref.\cite{Vergeles2019}
The similar scheme of symmetry breaking takes place for the spin-triplet $p$-wave superfluid $^3$He-B,\cite{VollhardtWolfle,Volovik1990} where instead of the Lorentz groups one has three-dimensional rotations in the orbital and spin spaces: $SO(3)_{L}\times SO(3)_{S}\rightarrow SO(3)_{S+L}$.

According to Eq.(\ref{bilinear}), the frame field $e_\mu^a$ transforms as a derivative in the same manner as the elasticity tetrads. That is why it has the dimension of inverse length, 
$[e_\mu^a]=1/[l]$, i.e. its dimension is shifted from $d=0$ to $d_{DV}=1$, see Eq.(\ref{fermions}).   In such vacua it is natural to assume that the fermionic field $\psi$ as well as the bosonic fields $\Phi$ are scalars under diffeomorphisms,\cite{Diakonov2011,VladimirovDiakonov2012}  i.e. their dimensions are shifted from $d=3/2$ and $d=1$ to $d_{\rm DV}=0$, see   Eqs.(\ref{fermions}) and (\ref{scalar_vector}) respectively.
For Weyl or massless Dirac fermions one has the conventional action:
\begin{equation}
S= \int d^4x |e| e^{a\mu}\left(\psi^\dagger \Gamma^a\nabla_\mu \psi + {\rm H.c.} \right)
\,.
\label{DiracZeroMass}
\end{equation}
The action (\ref{DiracZeroMass}), when expressed in terms of the Diakonov-Vladimirov (DV) tetrads, remains dimensionless, since 
$[e]=[l]^{-4}$, $[e^{a\mu}] =[l]$ and $[\psi]=1$, see  Eq.(\ref{fermions}). This suggests that the DV dimension of tetrads is natural, which is also supported by the elasticity tetrads.

The nontrivial dimension of the metric suggests that metric is not the quantity, which describes the space-time, but the quantity, which determines the dynamics of effective low energy fields in the background of microscopic quantum vacuum.

 \begin{equation}
\begin{matrix}
{\rm general \, relativity} & {\rm dimension} \,d & d_{\rm DV}\\
  &   & \\
g^{\mu\nu} & 0 & -2 \\
g_{\mu\nu} & 0 &  2 \\
\sqrt{-g} & 0 &  4 \\
d^4x \, \sqrt{-g} & -4 &  0 \\
ds^2 =g_{\mu\nu}  dx^\mu dx^\nu& -2 &  0 \\
dA=\sqrt{dS^{\mu\nu}dS_{\mu\nu}} & -2& 0 \\
M & 1 & 0 \\
S=M \int ds & 0 & 0 \\
\partial_{\mu} &1& 1 \\
p_{\mu} &1& 1 \\
R=g^{\mu\nu}R_{\mu\nu} & 2& 0 \\
G_{\rm Newton}  & -2& 0 \\
R/G_{\rm Newton}  &4 & 0 \\
\Lambda_{\rm cosmological} & 4& 0 \\
T_{\rm Hawking} & 1& 0 \\
T_{\rm Tolman} & 1& 1 \\
T_{\rm Tolman}/\sqrt{g_{00}} & 1& 0 \\
\end{matrix}
\label{GRdimensions}
\end{equation}

\begin{equation}
\begin{matrix}
{\rm scalar/vector}  & {\rm dimension}\,d & d_{\rm DV}\\
  &   & \\
\Phi & 1& 0\\
g^{\mu\nu}\partial_\mu \Phi \partial_\nu \Phi & 4 & 0 \\ 
M^2 \phi^2  & 4 & 0 \\  
A_{\mu} &1& 1  \\
F_{\mu\nu}& 2 & 2 \\ 
F_{\mu\nu}F^{\mu\nu}& 4 & 0\\ 
(F_{\mu\nu}F^{\mu\nu})^k& 4k & 0\\ 
F_{\mu\nu}\tilde F^{\mu\nu}& 4 & 4\\ 
F_{\mu\nu\alpha\beta}& 4& 4 \\ 
F^{\mu\nu\alpha\beta}& 4& -4 \\ 
q^2=F^{\mu\nu\alpha\beta}F_{\mu\nu\alpha\beta}& 8 & 0 \\ 
\mu_q& 0 & 0 \\ 
\end{matrix}
\label{scalar_vector}
\end{equation}

\begin{equation}
\begin{matrix}
{\rm fermions}  & {\rm dimension}\,d & d_{\rm DV}\\
  &   & \\
  e_\mu^a &0& 1\\
e^\mu_a &0& -1\\
e=\sqrt{-g}  &0& 4\\
\psi & \frac{3}{2} & 0\\
M\bar\psi \psi &4& 0\\
i\bar\psi \Gamma^a  e^\mu_a  D_\mu \psi &4& 0\\  
 \mathcal{T}_a &1& 2\\
  \mathcal{T}_a \mathcal{T}^a&2& 4\\
  \lambda_{\rm Nieh-Yan}^2&2& 0\\
   e^a_\mu A_\nu  \tilde F^{\mu\nu}&3& 4\\
  i e_a^\mu e_b^\nu \bar\psi    (\Gamma^a \Gamma^b - \Gamma^b \Gamma^a) \psi F_{\mu\nu}&5& 0\\
QQQL &6& 0\\
\end{matrix}
\label{fermions}
\end{equation}

\section{Physics in DV dimensions}
 \subsection{Mass terms}
 
The shifts of dimensions are shown in Eqs. (\ref{GRdimensions}),  (\ref{scalar_vector}) and  (\ref{fermions}) correspondingly  for gravity, scalar fields and fermions. Many quantities, which obey diffeomorphism invariance, become dimensionless. The action is dimensionless and remains dimensionless in the DV dimensions, since the action is diffeomorphism invariant.  Another example of the diffeomorphism invariant quantity is  the rest mass $M$ of particles. In the case of mass the dimension is shifted from $d=1$ to $d_{\rm DV}=0$.
That the DV dimension of mass is $[M]=1$ can be seen from classical equation for the particle spectrum: 
 \begin{equation}
g^{\mu\nu}p_\mu p_\nu = M^2
\,.
\label{MassDimension}
\end{equation}
According to the equation Eq.(\ref{GRdimensions}) for the shift of dimension, the Eq.(\ref{MassDimension}) gives for $M^2$ the dimension $d_{\rm DV}=-2+1+1=0$.  With dimensionless $[M]=1$ the classical action and the mass terms in the fermionic and bosonic actions remain dimensionless:
 \begin{eqnarray}
 S=M\int ds
\,,
\label{MassAction}
\\
S= \int d^4x |e| M\psi^\dagger \psi 
\,,
\label{MassTermFermion}
\\
S= \int d^4x\sqrt{-g} \left( g^{\mu\nu} \nabla_\mu \Phi  \nabla_\nu \Phi + M^2 \Phi^2 \right)\,,
\label{BosonicMass}
\\
S=
\frac{1}{4}\int d^4x \sqrt{-g} 
\left( F_{\mu\nu} F^{\mu\nu} + M^2 g^{\mu\nu}A_\mu A_\nu\right)
 \,.
\label{GaugeBosonAction}
\end{eqnarray}
This follows from the DV dimensions: $[e]=[\sqrt{-g}]=[l]^{-4}$, $[\Phi]=[\psi]=1$,  $[g^{\mu\nu}]=[l]^{-2}$, $[ds]=[M]=1$,  $[A_\mu]= [l]^{-1}$.

 \subsection{General relativity}
 
 Since the scalar curvature in general relativity is diffeomorphism invariant, it becomes dimensionless in the DV approach:  
\begin{equation}
[{\cal R}] =1\,.
\label{eq:R_dim}
\end{equation}
Its dimension is shifted from $d=2$ to $d_{\rm DV}=0$.
The other examples of the diffeomorphism invariant quantities are  the  Newton constant $G$ and the cosmological constant $\Lambda$ in the Einstein-Hilbert action:
\begin{equation}
S_{\rm{GR}}  = \frac{1}{16\pi G} \int d^4 x \sqrt{-g} {\cal R} + \int d^4 x \sqrt{-g}  \Lambda\,.
\label{GR}
\end{equation}
The dimensions of $G$ and $\Lambda$ are shifted from correspondingly $d=-2$ and $d=4$ to $d_{\rm DV}=0$, i.e.
 $[\Lambda] =[G]=1$. 
 These dimensionless quantitites are determined by the ratio of the mass scales \cite{Zeldovich1968} or by the functions of scalar fields.\cite{Starobinsky1980}
 In principle, only the ratio between the mass parameters makes sense.\cite{VladimirovDiakonov2012} In a given case only the combination $\Lambda G^2$ matters.
 According to Zeldovich,\cite{Zeldovich1968} this combination is expressed in terms of QCD mass scale:  $\Lambda G^2\sim \Lambda_{\rm QCD}^6 G^3$ (see also Refs. \cite{Schutzhold2002,KlinkhamerVolovik2009,UrbanZhitnitsky2010,BarvinskyZhitnitsky2018}). In the other approaches the electroweak energy scale \cite{ArkaniHamed2000,KlinkhamerVolovik2009b} and the neutrino mass scale\cite{KlinkhamerVolovik2011} may enter,  $\Lambda G^2\sim M_{W}^8 G^4$ and  $\Lambda G^2\sim M_{\rm n}^4 G^2$ respectively.
 
 On the other hand,  the spacetime volume $V=\int d^4x \sqrt{-g}$ is dimensionless, $[V]=1$, and thus in principle may have quantized values. Then it is not excluded that $\Lambda$, which in this case serves as the corresponding Lagrange multiplier, has universal quantized values, with  $\Lambda=0$ in the equilibrium Minkowski vacuum. 
 
 The dimensionless  Lagrange multiplier appears also in the $q$-theory of the quantum vacuum \cite{KlinkhamerVolovik2008}, if
the $q$-theory is based on the  $4$-form gauge field introduced by Hawking for phenomenological description of  the quantum vacuum,\cite{Hawking1984} $q^2=F^{\mu\nu\alpha\beta}F_{\mu\nu\alpha\beta}$. In the DV units both the vacuum variable $q$ and the Lagrange multiplier $\mu_q$ (the corresponding chemical potential of the conserved quantity) are dimensionless, see Eq.(\ref{scalar_vector}). The  Lagrange multiplier $\mu_q$ is dimensionless in both units. If it is fundamental, it becomes the general characteristics of the quantum vacuum. While the variable $q$ determines the variable cosmological constant $\Lambda(q)=\epsilon(q) - \mu_q q$, the universal chemical potential provides the nullification of $\Lambda$ in the Minkowski vacuum. At this fixed value of $\mu_q$ all the initial states (even those with the Planck scale 
$\Lambda$) finally relax to Minkowski vacuum with $\Lambda=0$, \cite{KlinkhamerVolovik2008} thus providing the solution of the cosmological constant problems.

\subsection{Mass, energy, temperature}

In the DV approach, mass and energy have different dimensions. While mass is dimensionless, $[M]=1$, the energy has dimension of frequency, 
$[E]=[\omega]=[\sqrt{g_{00}}]=1/[l]$. Correspondingly, the temperature is dimensionless, $[T]=1$, while the constant temperature, which enters the Tolman law:\cite{Tolman1934}
\begin{equation}
T({\bf r}) \sqrt{-g_{00}({\bf r})}= T_\text{Tolman}\,,
\label{eq:Tolman}
\end{equation}
has dimension of frequency, $[T_\text{Tolman}]=[\omega]=[\sqrt{g_{00}}]=1/[l]$, see 
Eq.(\ref{GRdimensions}).  Tolman temperature is the integration constant in equilibrium  in a stationary spacetime.\cite{Visser2019}

The Unruh temperature of the  accelerated  body is\cite{Unruh1976}
\begin{equation}
T_\mathrm{U}=\frac{ a}{2\pi}\,,
\label{eq:Unruh}
\end{equation}
where $a$ is covariant acceleration:
\begin{equation}
a^2 =
g_{\mu\nu}\frac{d^2 x^\mu}{ds^2}\frac{d^2 x^\nu}{ds^2} \,.
\label{eq:acceleration}
\end{equation}
Since $a$ is diffeomorphism-invariant, it is dimensionless together with the Unruh temperature,  $[a] = [T_\mathrm{U}]=1$. 

The same is with the Hawking temperature of a black hole. For the Schwartzschild black hole, with rest
energy $M_\mathrm{BH}$, Bekenstein entropy $S_\text{BH}$,  Hawking temperature $T_\text{BH}$  and horizon area $A_\text{BH}$,  one has:
 \begin{equation}
  T_\text{BH}=\frac{1}{8\pi GM_\mathrm{BH}}~~,~~S_\text{BH}=4\pi GM_\text{BH}^2 =\frac{A_\text{BH}}{4G}   \,.
\label{eq:HawkingT}
\end{equation}
All the quantities, that enter Eq.(\ref{eq:HawkingT}), are dimensionless in the DV approach, 
\begin{equation}
  [T_\text{BH}] =[S_\text{BH}] = [M_\text{BH}]=[A_\text{BH}]=[G]=1 \,.
\label{BHdimensions}
\end{equation}
The area of the black hole is dimensionless, because the covariant form of the scalar area element $dA$ is:
\begin{equation}
dA=\sqrt{dS^{\mu\nu}dS_{\mu\nu}} \,.
\label{CovariantArea}
\end{equation}
 Since  $[S^{\mu\nu}]=[l]^2$ and $[S_{\mu\nu}]=[S^{\mu\nu}][g_{\mu\nu}]^2=1/[l]^2$, one obtains $[A]=1$,
 which supports the idea that the area of the black hole horizon is quantized.\cite{Bekenstein1974,BekensteinMukhanov1995,Cardoso2019}
 
Similar quantization may occur for the de Sitter spacetime, which is the submanifold of Minkowski spacetime  in the 4+1 dimension:
 \begin{equation}
g_{\mu\nu}^{4+1\, {\rm Mink}}x^\mu x^\nu = \alpha^2 \,.
\label{dS}
\end{equation}
Since $[g_{\mu\nu}]=[l]^{-2}$, the parameter $\alpha$ is dimensionless, as well as the scalar curvature $R=12/\alpha^2$, i.e. in the DV dimensions $[R]=[\alpha]=1$.
The dimensionless parameter $\alpha$ of the de Sitter spacetime emphasizes the unique  symmetry of this spacetime and supports quantization of this parameter (see e.g. \cite{LopezOrtega2009}), which could be similar to the Bekenstein quantization of the black hole area.\cite{Bekenstein1974} However, in case of the superplastic vacuum the quantization of area can be very different from the quantization in terms of the Planck area, because the elementary cell of the underlying lattice may have nothing to do with the Planck scale.

 \subsection{Higher dimensional operators}

Eq.(\ref{fermions}) contains the operators with the mass dimensions 3, 5 and 6. The non-renormalisable dimension 5 operator gives a contribution to the electron magnetic moment.\cite{Bass2020},
\begin{equation}
G_5= i e_a^\mu e_b^\nu \bar\psi    (\Gamma^a \Gamma^b - \Gamma^b \Gamma^a) \psi F_{\mu\nu}\,,
\label{G5}
\end{equation}
and the non-renormalisable dimension 6  four-fermion operator  describes the baryon number violation:
\begin{equation}
G_6=QQQL \,,
\label{G6}
\end{equation}
where $L$ and $Q$ are the lepton and quark doublets.  Since in the DV approach the mass is dimensionless,
these operators become dimensionless: their $d_{DV}=0$. 
The prefactors in these terms are determined either by the ratio of the mass scales ("ultraviolet" and "infrared") or by the functions of scalar fields.
The same is with the $4k$ mass operator for $k>1$ in Eq.(\ref{scalar_vector}):
\begin{equation}
G_{4k}=(F_{\mu\nu}F^{\mu\nu})^k \,.
\label{G4k}
\end{equation}

 \section{Topological terms}

 \subsection{Dimension 4 operators}
 
 In terms of the DV dimensionalities, the  operators with $d_{DV}=4$ are topological. The operators of the type
 \begin{equation}
G_4= {\bf F}_{\mu\nu}\tilde{\bf F}^{\mu\nu}\,,
\label{G4}
\end{equation}
are topological in both classes of dimensions, since for them $d=d_{DV}=4$. 
They are accompanied by the fundamental integer of fractional prefactors.

But there are operators, which have original dimension $d\neq 4$ but acquire dimension $d_{DV}=4$ in the DV approach. This means that they are not topological in conventional approach, but may become topological in the DV dimensions.

The  former dimension $d=2$ operator $\mathcal{T}_a \mathcal{T}^a$ and the dimension $d=3$ operator $e^a_\mu A_\nu  \tilde F^{\mu\nu}$ acquire dimension $d_{DV}=4$ in  the DV dimensionalities. As a result they become topological and their prefactors become the topological quantum numbers. The operator $e^a_\mu A_\nu  \tilde F^{\mu\nu}$  determines the quantum Hall response in 3+1 topological insulators,\cite{NissinenVolovik2018,NissinenVolovik2019,Vishwanath2019} which we discuss in the Sec.\ref{HallEffect}.
The operator $\mathcal{T}_a \mathcal{T}^a$, where $\mathcal{T}_a$ is torsion, is discussed in Sec.\ref{NY}.

\subsection{3+1 intrinsic quantum Hall effect}
\label{HallEffect}

 The 3+1 topological insulator has the anomalous Hall response described by the following Chern-Simons topological term:\cite{NissinenVolovik2019}
\begin{eqnarray}
S_{4D}[A_{\mu}]=\frac{1}{4\pi^2} \sum_{a=1}^3N_a  
\int d^4 x~ E^{~a}_{\mu} \epsilon^{\mu\nu\alpha\beta} A_\nu \partial_\alpha A_\beta\,.
\label{action}
\end{eqnarray}
It explicitly contains the elasticity tetrads $E^{~a}_{\mu}$, which is full derivative in the absence of dislocations. The  integer coefficients $N_a$ are three topological invariants, the winding numbers, expressed in terms of integrals of the Green's functions in the energy-momentum space:
\begin{eqnarray}
N_a=\frac{1}{8\pi^2}\epsilon_{ijk} \int_{-\infty}^{\infty} d\omega\int_{\rm BZ} dS_a^i
\nonumber
\\
{\rm Tr} [(G_{\omega} G^{-1}) (G_{k_i} G^{-1}) ( G_{k_j} G^{-1})]\,.
\label{Invariants}
\end{eqnarray}
The momentum integral is over the 2D torus in the cross section $\bf{S}_a$ of the three-dimensional Brillouin zone.  
The gauge invariance of the action (\ref{action})  is supported by the condition (\ref{integrability}) even in the presence of deformations.
The topological invariants $N_a$ describe the quantized  response of the Hall conductivity to the deformation of the crystal:
\begin{equation}
\frac{d\sigma_{ij}}{dE^{\ a}_k} =\frac{e^2}{2\pi h}\epsilon_{ijk}  N_a\,.
\label{ConductivityVariation}
\end{equation}

Discussions of  the 3+1 quantum Hall effect in Weyl semimetals can be found in Ref.\cite{Burkov2020}

\subsection{Nieh-Yan anomaly}
\label{NY}

Another example is the Nieh-Yan anomaly related to torsion.
\cite{NiehYan1982a,NiehYan1982b,Nieh2007,Yajima96,ChandiaZanelli97a,ChandiaZanelli97b,ChandiaZanelli98,Obukhov1997,Parrikar2014, FerreirosEtAl19,Nissinen2020}
 For the conventional torsion and curvature in terms of the conventional dimensionless tetrads, the gravitational Nieh-Yan anomaly equation for the non-conservation of the axial current
\begin{equation}
\partial_\mu   j_5^\mu =  \lambda^2
\left(\mathcal{T}^a \wedge \mathcal{T}_a - e^a \wedge e^b\wedge R_{ab} \right) \,, 
\label{eq:NYterm}
\end{equation}
contains the nonuniversal prefactor -- the ultraviolet cut-off parameter $\lambda$ with dimension of inverse length, $[\lambda]=1/[l]$. Because of such prefactor,  the Nie-Yan contribution to the anomaly is still rather subtle (see recent literature 
\cite{Khaidukov2018,Nissinen2020,NissinenVolovik2019,Stone2019,Ojanen2020,Stone2020}). This is  because the nonuniversal parameter may depend on the spacetime coordinates, which explicitly violates the topology.

However, in terms of DV tetrads, the torsion in Eq.(\ref{Torsion}) has dimension $\left[T^a_{kl}\right] =1/[l]^2$, and the Nieh-Yan operator 
has the proper dimension $d_{\rm DV}=4$. The prefactor $\lambda^2$ of the topological operator becomes dimensionless, 
$[\lambda]=1$, which suggests that the prefactor is universal and is quantized.

\subsection{Wess-Zumino action}

The Chern-Simons term in Sec.\ref{HallEffect},  which describes the  3+1 QHE, can be extended to the 3+1+1 Wess-Zumino action:
\begin{equation}
S_{\rm{WZ}}^{a{\rm a}{\rm b}} = \frac{1}{8\pi^2}\int_{X^5} d^4x d\tau\,   \epsilon^{\mu\nu \alpha \beta\gamma} 
e^{~a}_{\gamma}F^{\rm a}_{\mu\nu} F^{\rm b}_{\alpha \beta}\,.
\label{WessZuminoEFF}
\end{equation}
In the DV dimensions it is topological, and it can be extended to the Nieh-Yan anomaly in terms of torsion:
 \begin{equation}
S_{\rm{NY}}^{abc} \propto \int_{X^5} d^4x d\tau\,   \epsilon^{\mu\nu \alpha \beta\gamma} 
e^{~a}_{\gamma}T_{\mu\nu}^b T_{\alpha \beta}^c \,.
\label{NiehYan5D}
\end{equation}

There is a set of the mixed Wess-Zumino terms, which contain both the torsion and gauge field, such as:
\begin{equation}
S_{\rm{WZ}}^{ab} \propto\int_{X^5} d^4x d\tau\,   \epsilon^{\mu\nu \alpha \beta\gamma} 
e^{~a}_{\gamma}T^b_{\mu\nu} F_{\alpha \beta}\,.
\label{WessZuminoETF}
\end{equation}
In terms of DV tetrads, these dimensionless terms are also universal and do not depend on the cut-off parameters.

\section{Conclusion}

In two scenarios of emergent gravity, the superplastic vacuum and the DV theory with bilinear tetrad field, the invariance under diffeomorphisms leads to the dimensionless physics.   In the DV  theory this invariance is assumed as fundamental.  In the superplastic vacuum, the diffeomorphism invariance corresponds to the proposed invariance under arbitrary deformations of the 4D vacuum crystal.
In words of 't Hooft (applied originally to the local conformal symmetry) "this could be a way to make distance and time scales relative, so that what was dubbed as ‘small distances’ ceases to have an absolute meaning."\cite{tHooft2015}  All this may suggest that the dimensionless physics can be the natural consequence of the diffeomorphism invariance, and thus can be the general property of any gravity, which emerges in the quantum vacuum.

The dimensionless physics emerging in the frame of the DV dimensionful tetrads leads in particular to the new topological terms in action, since some of the dimensionless parameters appear  to be the integer valued quantum numbers, which characterize the topology of the quantum vacuum. This has been seen on example of the 3+1 dimensional quantum Hall effect in topological insulators. When the Chern-Simons action is written in terms of  the elasticity or DV tetrads, its prefactor becomes dimensionless and universal, being expressed in terms of  integer-valued momentum-space invariant. 

Another example is the phenomenon of the chiral anomaly in terms of the torsion fields suggested by Nieh and Yan. For the conventional torsion and curvature in terms of the conventional dimensionless tetrads, the gravitational Nieh-Yan anomaly equation for the non-conservation of the axial current contains the nonuniversal prefactor -- the ultraviolet cut-off parameter $\lambda$ with dimension of inverse length. This parameter  may depend on the spacetime coordinates, which explicitly violates the topology.
However, in terms of DV tetrads, the dimension of torsion is shifted form $d=1$ to $d_{\rm DV}=2$. As a result the prefactor 
$\lambda^2$ becomes dimensionless and thus becomes universal. 

However, the universality takes place only for the topological numbers and the symmetry parameters. The other dimensionless quantitites are not universal, being described by the functions of the ratios of different mass scales. In this respect the answer to the question of how many fundamental constants there are in physics\cite{Trialogue,Uzan2003,Uzan2011,Duff2015} can be trivial:
there are no  fundamental constants, and the ratios of parameters and the ratio of the length scales are the only meaningful quantities.\cite{VladimirovDiakonov2012}

{\bf Acknowledgements}.  I thank Friedrich Hehl and Yuri Obukhov for sending me their papers. This work has been supported by the European Research Council (ERC) under the European Union's Horizon 2020 research and innovation programme (Grant Agreement No. 694248).


\begin{thebibliography}{99}


\bibitem{Bass2020}
S.D. Bass,
Emergent gauge symmetries and particle physics,
Progress in Particle and Nuclear Physics {\bf 113},  103756 (2020).

 \bibitem{Sakharov}
A.D. Sakharov, Vacuum quantum fluctuations in
curved space and the
theory of gravitation, Dokl. Akad. Nauk {\bf 177}, 70--71 (1967) [Sov. Phys.
Dokl. {\bf 12}, 1040--41 (1968)]; reprinted in Gen. Relative. Gravity {\bf
32}, 365--367 (2000).

\bibitem{Unruh1981} 
W.G. Unruh,  
Experimental black-hole evaporation?
Phys. Rev. Lett. {\bf 46}, 1351 (1981).

\bibitem{Nielsen1976}
H.B. Nielsen,
 “Dual Strings” - Section 6. Catastrophe Theory Programme,
  in: I.M. Barbour \& A.T. Davies (eds.), “Fundamentals of Quark Models”, Scottish Univ. Summer School in Phys. (1976) pp. 528-543.

\bibitem{Volovik1986} 
G.E. Volovik, 
Analog of gravity in superfluid  $^3$He-A,
Pisma ZhETF {\bf 44}, 388--390 (1986);  JETP Lett. {\bf 44}, 498--501 (1986).

\bibitem{Froggatt1991}
C.D. Froggatt   and  H.B. Nielsen,
{\it Origin of Symmetry}, 
World Scientific, Singapore, 1991.

\bibitem{Horava2005}  
P. Ho\v{r}ava,
Stability of Fermi surfaces and $K$-theory,
Phys. Rev. Lett. \textbf{95}, 016405 (2005).

\bibitem{Volovik2003} 
G.E. Volovik, 
{\it The Universe in a Helium Droplet}, 
Clarendon Press,  Oxford (2003).

 \bibitem{Diakonov2011}
 D. Diakonov,
 Towards lattice-regularized Quantum Gravity,
 arXiv:1109.0091.

 \bibitem{VladimirovDiakonov2012}
A.A. Vladimirov and D. Diakonov,
Phase transitions in spinor quantum gravity on a lattice,
Phys. Rev. D {\bf 86}, 104019 (2012).

 \bibitem{VladimirovDiakonov2014}
A.A. Vladimirov and D. Diakonov,
Diffeomorphism-invariant lattice actions,
Physics of Particles and Nuclei {\bf 45}, 800 (2014).

 \bibitem{ObukhovHehl2012}
Y.N. Obukhov and F.W. Hehl,
Extended Einstein–Cartan theory a la Diakonov: The field equations,
Phys. Lett. B {\bf 713}, 321--325 (2012).

\bibitem{Volovik1990}
G. E. Volovik, 
Superfluid $^3$He-B and gravity,
Physica (Amsterdam) {\bf 162}B, 222--230 (1990).

\bibitem{Bilby1955} 
 B.A. Bilby and E. Smith,
 Continuous distributions of dislocations: A new application of the methods of non-Riemannian geometry, 
 Proc. Roy. Soc. Sect. A {\bf 231}, 263--273 (1955).
 
 \bibitem{Bilby1956} 
 B.A. Bilby and E. Smith, 
Continuous distributions of dislocations. III, 
 Proc. Roy. Soc. Sect. A {\bf 236}, 481--505 (1956).
 
 \bibitem{Kroener1960}
 E. Kr\"oner,
Allgemeine Kontinuumstheorie der Versetzongen and Ligenspannunge,
 Arch. Rational Mech. Anal. {\bf 4}, 18--334 (1960).

\bibitem{DzyalVol1980}
I.E. Dzyaloshinskii, and G.E. Volovick, 
Poisson brackets in  condensed matter,
Ann. Phys.  {\bf 125} 67--97 (1980).

\bibitem{VolovikDotsenko1979}
G.E. Volovik, V.S. Dotsenko (jr), 
Poisson brackets and continual dynamics of the vortex lattice in rotating HeII,
Pisma ZhETF {\bf 29}, 630--633 (1979);  JETP Lett. {\bf 29} 576--579 (1979).

\bibitem{AndreevKagan84}
A.F. Andreev, M. Yu. Kagan,
Hydrodynamics of a rotating superfluid liquid,
Zh. Eksp. Teor. Fiz. {\bf 86},546 (1984) [Sov. Phys. JETP {\bf 59} (1984)]

\bibitem{KleinertZaanen2004} 
H. Kleinert and J. Zaanen, 
World nematic crystal model of gravity explaining the absence of torsion, 
Phys. Lett. A {\bf 324}, 361--365 (2004).

\bibitem{HehlObukhov2007} 
F.W. Hehl and Y.N. Obukhov,
Elie Cartan’s torsion in geometry and in field theory, an essay,
Annales de la Fondation Louis de Broglie {\bf 32}, 157 --194 (2007).

 \bibitem{KlinkhamerVolovik2019}
F.R. Klinkhamer and G.E. Volovik,
Tetrads and $q$-theory,
Pis'ma ZhETF  {\bf 109}, 369--370 (2019),
 JETP Lett. {\bf 109},  362--365 (2019),
arXiv:1812.07046.

 \bibitem{Zubkov2019}
M.A. Zubkov,
Emergent gravity in superplastic crystals and cosmological constant problem,
arXiv:1909.08412 [gr-qc].

 \bibitem{Andreev1969}
A.F. Andreev, I.M. Lifshitz,
Quantum theory of defects in crystals, 
ZhETF {\bf 56}, 2057 (1969),
JETP {\bf 29}, 1107 (1969).

 \bibitem{Dzyaloshinskii1972}
I.E. Dzyaloshinskii, P.S. Kondratenko, V.S. Levchenkov,
Theory of quantum crystals. I. Phenomenological theory,
ZhETF {\bf 62}, 1574--1584 (1972), 
 JETP {\bf 35}, 823--828 (1972).

 \bibitem{Dzyaloshinskii1972b}
I.E. Dzyaloshinskii, P.S. Kondratenko, V.S. Levchenkov, 
ZhETF {\bf 62}, 2318--2132 (1972), JETP {\bf 35}, 1213--1219 (1972).

 \bibitem{NissinenVolovik2018}
J. Nissinen and G.E. Volovik,
Tetrads in solids: from elasticity theory to topological quantum Hall systems and Weyl fermions,
ZhETF {\bf 154},   1051 (2018),
JETP {\bf 127}, 948 (2018),
arXiv:1803.09234.

 \bibitem{NissinenVolovik2019}
J. Nissinen and G.E. Volovik,
Elasticity tetrads, mixed axial-gravitational anomalies, and (3+1)-d quantum Hall effect,
PRResearch {\bf 1}, 023007 (2019),
arXiv:1812.03175.

 \bibitem{Volovik2020}
G.E. Volovik,
On dimension of tetrads in effective gravity,
Pis’ma v ZhETF {\bf 111},  411--412 (2020),
JETP Lett. {\bf 111}, 368--370 (2020),
arXiv:2003.00915.

 \bibitem{Akama1978}
K. Akama,
An attempt at pregeometry: gravity with composite metric, 
Progress of Theoretical Physics {\bf 60}, 1900--1909 (1978).

\bibitem{Vergeles2019}
S.N. Vergeles,
A note on the vacuum structure of lattice Euclidean quantum gravity: birth of macroscopic space-time and $PT$-symmetry breaking,
arXiv:1903.09957.

 \bibitem{VollhardtWolfle}
D. Vollhardt and P. Wölfle, 
The Superfluid Phases of Helium 3 (Taylor \& Francis, London, 1990).

\bibitem{Zeldovich1968}
Ya.B. Zel'dovich,
The Cosmological constant and the theory of elementary particles,
Sov. Phys. Usp. {\bf 11}, 381--393  (1968).

\bibitem{Starobinsky1980}
A.A. Starobinsky,
A new type of isotropic cosmological models without singularity, 
Phys. Lett. B {\bf 9}, 99 (1980).

\bibitem{Schutzhold2002}
R. Sch\"utzhold,
Small Cosmological Constant from the QCD Trace Anomaly?
Phys. Rev. Lett. {\bf 89}, 081302 (2002),

\bibitem{KlinkhamerVolovik2009}
F.R. Klinkhamer and G.E. Volovik,  
Gluonic vacuum, $q$-theory, and the cosmological constant, 
Phys. Rev. D {\bf 79}, 063527 (2009).

\bibitem{UrbanZhitnitsky2010}
F.R. Urban and A.R. Zhitnitsky,
The QCD nature of dark energy,
Nucl. Phys. B {\bf 835},  135--173 (2010).

\bibitem{BarvinskyZhitnitsky2018}
A.O. Barvinsky and A.R. Zhitnitsky,
 Inflation and gauge field holonomy,
 Phys. Rev. D {\bf 98}, 045008 (2018).

\bibitem{ArkaniHamed2000}
N. Arkani-Hamed, L. J. Hall, C. Kolda, and H. Murayama,
New perspective on cosmic coincidence problems,
Phys. Rev. Lett. {\bf 85}, 4434 (2000).

\bibitem{KlinkhamerVolovik2009b}
F.R. Klinkhamer and G.E. Volovik, 
Vacuum energy density kicked by the electroweak crossover,
Phys. Rev.  D {\bf 80}, 083001 (2009); 
arXiv:0905.1919.

\bibitem{KlinkhamerVolovik2011}
F.R. Klinkhamer and G.E.Volovik,
Dynamics of the quantum vacuum: Cosmology as relaxation to the equilibrium state,
Journal of Physics: Conference Series {\bf 314}, 012004 (2011);
arXiv:1102.3152.

\bibitem{KlinkhamerVolovik2008}
F.R. Klinkhamer and G.E. Volovik,  
Dynamic vacuum variable and equilibrium approach in cosmology,  
 Phys. Rev. D {\bf 78}, 063528 (2008); 
arXiv:0806.2805 [gr-qc].

\bibitem{Hawking1984}
S.W. Hawking,
The cosmological constant is probably zero,
Phys. Lett.  B {\bf 134}, 403 (1984).


\bibitem{Tolman1934}
R.C. Tolman,
 {\it Relativity, Thermodynamics and Cosmology},
 Clarendon Press, Oxford (1934).

\bibitem{Visser2019}
J. Santiago and M. Visser,
Tolman temperature gradients in a gravitational field,
Eur. J. Phys. {\bf 40}, 025604  (2019).

\bibitem{Unruh1976}
W.G. Unruh,
Notes on black-hole evaporation,
Phys. Rev. D {\bf 14}, 870 (1976).

\bibitem{Bekenstein1974}
 J. D. Bekenstein, 
 The quantum mass spectrum of the Kerr black hole,
 Lett. Nuovo Cimento {\bf 11}, 467 (1974).

\bibitem{BekensteinMukhanov1995}
J.D. Bekenstein and V.F. Mukhanov, 
Spectroscopy of the quantum black hole, 
Phys. Lett. B {\bf 360}, 7 (1995), 
gr-qc/9505012.

\bibitem{Cardoso2019}
V. Cardoso, V.F. Foitc and M. Kleban,
Gravitational wave echoes from black hole area quantization,
JCAP08(2019)006.

\bibitem{LopezOrtega2009}
A. Lopez-Ortega,
Area spectrum of the D-dimensional de Sitter spacetime,
Physics Letters B {\bf 682}, 85--88 (2009).
 
\bibitem{Vishwanath2019}
Xue-Yang Song, Yin-Chen He, Ashvin Vishwanath, Chong Wang,
Electric polarization as a nonquantized topological response and boundary Luttinger theorem,
arXiv:1909.08637.

\bibitem{Burkov2020}
M. Thakurathi and A.A. Burkov,
Theory of the fractional quantum Hall effect in Weyl semimetals,
Phys. Rev. B {\bf 101}, 235168 (2020).

\bibitem{NiehYan1982a}
H.T. Nieh and M.L. Yan, 
An identity in Riemann-Cartan geometry,
 J. Math. Phys. {\bf 23}, 373  (1982).

 \bibitem{NiehYan1982b}
H.T. Nieh and M.L. Yan, 
Quantized Dirac field in curved Riemann-Cartan background. I. Symmetry properties, Green's function, 
Ann. Phys. {\bf 138}, 237 (1982).

\bibitem{Nieh2007}
H.T. Nieh,
A torsional topological invariant, 
Int. J. Mod. Phys. A {\bf 22},  5237 (2007).

\bibitem{Yajima96}
S. Yajima, 
Evaluation of the heat kernel in Riemann - Cartan space,
Class. Quantum Grav. {\bf 13}, 2423 (1996).

\bibitem{ChandiaZanelli97a}
O. Chandia and J. Zanelli, 
Topological invariants, instantons, and the chiral anomaly on spaces with torsion,
Phys. Rev. D {\bf 55}, 7580 (1997).
 
\bibitem{ChandiaZanelli97b}
O. Chandia and J. Zanelli, 
Torsional topological invariants (and their relevance for real life), 
arXiv:hep-th/9708139.

\bibitem{ChandiaZanelli98}
O. Chandia and J. Zanelli, 
Supersymmetric particle in a spacetime with torsion and the index theorem,
Phys. Rev. D 58, 045014 (1998).

\bibitem{Obukhov1997}
Y.N. Obukhov,  E.W. Mielke, J. Budczies and F.W. Hehl,
On the chiral anomaly in non-Riemannian spacetimes,
Foundations of Physics, {\bf 27}, 1221--1236 (1997).

 \bibitem{Parrikar2014}
O. Parrikar, T.L. Hughes, R.G. Leigh,
Torsion, parity-odd response and anomalies in topological states,
Phys. Rev. D {\bf 90}, 105004 (2014).

\bibitem{FerreirosEtAl19}
Y. Ferreiros, Y. Kedem, E.J. Bergholtz, and J.H. Bardarson,
Mixed axial-torsional anomaly in Weyl semimetals,
Phys. Rev. Lett. {\bf 122}, 056601 (2019).

\bibitem{Nissinen2020} 
J. Nissinen,
Emergent spacetime and gravitational Nieh-Yan anomaly in chiral $p + ip$ Weyl
superfluids and superconductors, 
Phys. Rev. Lett. {\bf 124}, 117002 (2020).
arXiv:1909.05846.

\bibitem{Khaidukov2018} 
Z.V. Khaidukov and M.A. Zubkov,
Chiral torsional effect,
Pis’ma v ZhETF, {\bf 108},  702--703 (2018) [JETP Lett. 108, 670 (2018)],
arXiv:1812.00970.

\bibitem{Stone2019} 
Ze-Min Huang, Bo Han and M. Stone,
The Nieh-Yan anomaly: torsional Landau levels, central charge and anomalous
thermal Hall effect, 
Phys. Rev. B {\bf 101}, 125201 (2020).
arXiv:1911.00174 (2019).

\bibitem{Ojanen2020} 
 Long Liang and T. Ojanen,
Topological magnetotorsional effect in Weyl semimetals,
Phys. Rev. Research {\bf 2}, 022016(R) (2020).

\bibitem{Stone2020} 
Ze-Min Huang, Bo Han and M. Stone,
Hamiltonian approach to the torsional anomalies and its dimensional ladder,
Phys. Rev. B {\bf 101}, 165201 (2020),
arXiv:1912.06051.

\bibitem{tHooft2015} 
G. 't Hooft,
{\it The cellular automaton interpretation of quantum mechanics},
 Fundamental Theories of Physics  {\bf 185},
arXiv:1405.1548v3.

\bibitem{Trialogue} M.J. Duff, L.B. Okun, and G. Veneziano,
Trialogue on the number of fundamental constants, 
JHEP {\bf 03023}, 1-30 (2002), physics/0110060.

\bibitem{Uzan2003} 
J.-P. Uzan,
The fundamental constants and their variation: observational and theoretical status,
Rev. Mod. Phys. {\bf 75}, 403 (2003).

\bibitem{Uzan2011} 
J.-P. Uzan,
Varying constants, gravitation and cosmology,
Living Rev. Relativ.  {\bf 14}, 2 (2011).  

\bibitem{Duff2015} 
 M.J. Duff, 
 How fundamental are fundamental constants?  
 Contemporary Physics {\bf 56},  35--47 (2015).
 
 \end{thebibliography}
\end{document}